# PiBase: An IoT-based Security System Using Google Firebase and Raspberry Pi

Venkat Margapuri
Department of Computer Science
Kansas State University
Manhattan, KS
marven@ksu.edu

Niketa Penumajji
Department of Computer Science
Kansas State University
Manhattan, KS
niketa912@ksu.edu

Mitchell Neilsen
Department of Computer Science
Kansas State University
Manhattan, KS
neilsen@ksu.edu

*Abstract*— Smart environments are environments where digital devices are connected to each other over the Internet and operate in sync. Security is of paramount importance in such environments. This paper addresses aspects of authorized access and intruder detection for "smart" environments. Proposed is PiBase, an Internet of Things (IoT)-based app that aids in detecting intruders and providing security. The hardware for the application consists of a Raspberry Pi, a PIR motion sensor to detect motion from infrared radiation in the environment, an Android mobile phone and a camera. The software for the application is written in Java, Python and NodeJS. The PIR sensor and Pi camera module connected to the Raspberry Pi aid in detecting human intrusion. Machine learning algorithms, namely Haar-feature based cascade classifiers and Linear Binary Pattern Histograms (LBPH), are used for face detection and face recognition, respectively. The app lets the user create a list of non-intruders and anyone that's not on the list is identified as an intruder. The app alerts the user only in the event of an intrusion by using the Google Firebase's Cloud Messaging service to trigger a notification to the app. The user may choose to add the detected intruder to the list of non-intruders through the app to avoid further detections as intruder. Face detection by the Haar Cascade algorithm yields a recall of 94.6%. Thus, the system is both highly effective and relatively low cost.

## I. Introduction

One of the major concerns for homes and businesses in the modern world is the threat of property intrusion and burglary. There are 2.5 million burglaries in the US annually according to the US Department of Justice yearly data and over half of them are home invasions [19]. An increased number of people in the work-force limits the amount of time they spend at home leaving homes vulnerable. Besides, the rise of e-commerce during the COVID-19 pandemic has led to a substantial rise in products shipped to homes and establishments. It is reported that 35.5 million people in the US alone have been victims of porch pirating with the average stolen package costing $156 [9]. The need to safeguard homes and establishments is paramount and security solutions are the need of the hour. The article presents one such solution using the notion of Internet of Things (IoT). IoT may be perceived of as the network of connected devices that communicate over the Internet while following a certain set of protocols to achieve a common goal. In other words, individual "smart" entities connected over the Internet form a "smart" network and operate in harmony.

Security cameras act as a formidable means to prevent and provide evidence of property intrusion after the fact. Generally, city dwellers place security camera systems that cost anywhere between a few hundred to a few thousand dollars to safeguard their homes. This article proposes PiBase, a low-cost, machine-to-machine (M2M) IoT application that aids in securing homes and establishments using machine learning. The application can also be used in other smart environments where authorized access is required. The ecosystem of the application consists of a Raspberry Pi, a Passive Infrared (PIR) motion sensor, a Raspberry Pi camera module and an Android mobile phone. The software for the system is developed in Java, Python and NodeJS. The user information is stored in Google Firebase, a cloud environment provided by Google. The use of Google's software solution ensures that user information is protected at all times and the information is robust to attacks compared with other home-grown solutions.

In the remainder of the article, Section II describes related work, Section III describes the hardware and software components, Section IV, the architecture and implementation and Section V sheds light on future work and concludes the article.

## II. Related Work

Chowdhury, Nooman and Sarker [5] proposed an access control system using Raspberry Pi for home security. The application is setup to send an email and tweet to the user anytime a face is detected at the door. The user is required to re-tweet with the door lock or open command to trigger the magnetic lock which in turn opens or keeps the door locked.

Pavithra and Balakrishnan [14] developed an implementation of IoT to monitor and control home appliances using a Raspberry Pi, a PIR sensor and a smartphone via the Internet. Automated control of lights, fans and fire detection in the event of a fire are demonstrated in the work.

Sajjad et al. [18] devised a face recognition framework using Raspberry Pi for enhanced law-enforcement services. The proposed approach used Viola Jones approach for face detection. Besides, the system used Bag of Words technique for extraction of oriented FAST and rotated BRIEF points from the detected face, followed by support vector machine for identification of suspects.

Prathaban, Thean and Sazali [15] developed a vision-based home security system using Raspberry Pi and OpenCV. The proposed solution leveraged Haar-Cascade algorithm coupled with background subtraction and Histogram of Oriented Gradients (HOG). The solution was tested against a system that only contained PIR sensors. It was observed that the OpenCV solution had an accuracy of 100% as opposed to the PIR sensor system's 76%.

Dash and Nayak [13] implemented a smart surveillance monitoring system using Raspberry Pi and PIR sensor for mobile devices. Notifications to the user's smartphone are sent over the internet upon the detection of motion to alert the user.

Nadaf, Hatture, Bonal and Naik [12] proposed a system designed as a smart mirror that provides both home security and other information such as news, weather information and calendar. The application leveraged the Amazon cloud to store the images captured and SMLLane servers to send SMS notifications to the user.

Ansari, Sedky, Sharma and Tyagi [3] developed a security system using Raspberry Pi, camera and PIR motion sensor to monitor and trigger alarms when motion is detected. The captured photos and videos are sent directly to a remote cloud server. In the event the cloud server is unavailable, they are stored on the Raspberry Pi temporarily until the cloud server becomes available.

Abaya, Basa, Sy, Abad and Dadios [1] designed a security system using Raspberry Pi and OpenCV. The developed system was put to test inside a warehouse facility. In the absence of a dedicated motion detection sensor, the system used background subtraction algorithm for motion detection. Detection for humans was performed using Haar features and triggered e-mail notification, an LED indicator and sound alarm. In the event the detected motion was not a human, the system checked for smoke by using the property, 'the direction of hot smoke is always upward in the absence of wind'.

Gupta, Patil, Kadam and Dumbre [7] explored the feasibility of implementing Raspberry Pi face detection and recognition using conventional face detection and recognition techniques such as Haar detection and PCA. The objective of the work was to propose a solution that could replace the use of passwords and RF I-Cards for access to high security systems and buildings besides lowering the investment in them.

Patil and Shukla [16] implemented a classroom attendance system based on face recognition in class using Raspberry Pi. The Viola Jones algorithm was used for face detection and face recognition using Principal Component Analysis (PCA) and Linear Discriminant Analysis (LDA).

### III. HARDWARE AND SOFTWARE COMPONENTS

The hardware components include a Raspberry Pi 3.0, PIR motion sensor, Pi Camera module and Android smartphone. The components are briefly described below.

*A. Raspberry Pi 3.0*

The Raspberry Pi 3.0 is a small-sized computer that has the ability to plug into a computer monitor or any other display using HDMI. It runs the Raspbian OS and is handy to run applications in programming languages such as Python. The Raspberry Pi is provided with a 40-pin General Purpose Input/ Output (GPIO) that helps to interact with external components such as the PIR motion sensor. The voltages and the input/ output capabilities of the pins are as follows:

- Voltages: Two 5V pins and two 3V pins are present on the Raspberry Pi along with a number of ground pins which are not configurable. The remainder of the pins are general purpose 3V pins meaning inputs are 3V tolerant and outputs are 3V.
- Output: A GPIO pin designated as output pin can be set to high (3V) or low (0V).
- Input: A GPIO pin designated as an input pin can be set to high (3V) or low (0V). Pull-up and pull-down resistors are used to do so. Only pins GPIO2 and GPIO3 have fixed pull-up resistors but the others may be configured in software.

*B. PIR Motion Sensor*

PIR motion sensors work on the principle of Pyroelectricity i.e. the property of materials where they generate a certain amount of voltage when subject to temperature changes. The PIR sensor consists of two slots, each one made of a material that is sensitive to infrared (IR) radiation and lens. Both slots have the ability to detect motion within the sensitivity range of the sensors. In the idle state where the sensors are exposed to ambient conditions, either slot detects the same amount of IR radiation. However, when an object approaches the sensor, it intercepts one half of the PIR sensor that causes a positive differential change between the two halves. As the object leaves the sensing area, the opposite happens and the sensor generates a negative differential change. The positive and negative differential changes trigger an output pulse of 3V. The lens around the sensors is a Fresnel lens. The use of Fresnel lens reduces the amount of material required compared to a conventional lens since it divides the lens into a set of concentric annular sections.

*C. Pi Camera Module*

The Pi Camera module is the camera provided for the Raspberry Pi by the Raspberry Pi Foundation. The camera used as part of the work is Camera Module v2, an 8-megapixel camera first released in 2016.

The software developed for the application is written in Java, Python and NodeJS. Each of the software components is as described further.

*A. Android Application*

The Android application acts as the UI for PiBase and is implemented in Java. The app provides the ability for the user to register, login, upload a list of intruders and view any intrusions that might occur. Google Firebase is used to authenticate and store any information entered by the user on the app such as the list of intruders. The notifications of intrusion which include a picture and timestamp are are received on the app. Initially, the notifications are displayed on Android's notification tray but the user may click on the notification to view it in the app.

*B. Google Firebase*

One of the cloud services provided by Google is Firebase. The environment provides a multitude of services such as user authentication and security, cloud to client messaging, database, emulator suite, hosting for micro-services, storage for user-generated content such as images and videos and plenty of other features. The ones pertinent to the current work are user authentication, database and cloud messaging.

*User Authentication:* Firebase provides a complete backend solution for authentication to sign-in with passwords, federated identity providers, email links and text messages. The Admin SDK shipped as part of Firebase helps to integrate authentication services into custom apps built in Android (or other platforms). The user may read and write data to the real-time database, send cloud messages, generate and verify Firebase auth tokens using the Admin SDK.

*Database:* Firebase provides users with a NoSQL real-time database where data is synced across all clients in real-time, and remains available at all times. The data in the database is stored in JSON. The clients are able to use references to the JSON data to write data and subscribe to data changes made by other clients attached to the database. The database also provides users with the ability to write security rules that regulate access to the data

*Storage:* Firebase Storage provides users with the ability to store and serve user-generated content, such as images or videos. The files are stored in a Google Cloud Storage bucket that makes them accessible through Firebase and Google Cloud. Cloud Storage SDK is available to access and manipulate the stored data. The SDK integrates with Firebase Authentication as well making it easier to setup user-based/group-based security.

*Firebase Cloud Messaging (FCM):* FCM is a cross-platform cloud-messaging service that lets users send messages from Firebase to clients. FCM is able to send messages to an individual client app or place messages on a 'topic'. A topic may be understood as a shared communication channel wherein the messages placed in a topic are available to all the clients that are subscribed to the topic, essentially enabling a publish-subscribe mechanism. The messages sent by FCM are classified into two categories: Notification and Data. The primary difference between the two categories is that Notification messages are displayed automatically on the user's console, be it a smartphone or tablet whereas Data messages are required to be processed by the client application. Besides, Notification messages have a predefined set of user-visible keys and optional data payload of custom key-value pairs whereas Data messages only contain custom key-value pairs with no reserved key names. Both message types have a maximum payload limit of 4000 bytes. The Admin SDK may be used to integrate and interact with FCM.

### C. Face Detection and Face Recognition

Face detection and recognition are performed using OpenCV, an open-source computer vision library. Face detection is performed using the Haar Cascades algorithm and recognition is performed using Linear Binary Pattern Histograms (LBPH) algorithm. Both algorithms are popular in the area of computer vision and are briefly described.

*Haar Cascades:* Haar Cascades, otherwise known as the Viola-Jones algorithm is an object detection algorithm to identify faces in images and videos. The algorithm leverages the proposed Haar Features that aid in detecting edges or lines on the images, or to identify areas on the image with varying intensities of pixels. The exhaustive set of features proposed by the Viola Jones algorithm contains 180,000 features. However, not every feature holds relevance to the task in question. Adaboost is a feature selection technique that helps to constrict the set of features. Adaboost applies each of the 180,000 features to the images in the dataset separately and creates weak learners. Weak learners are defined as classifiers that perform better than random guessing but still exhibit performance that is below-par in terms of classifying positive and negative images. The resulting feature set consists of fewer features eliminating the ones that are irrelevant. Finally, a technique known as Attentional Cascade is applied on each of the images to detect faces. In Attentional Cascade, the features are applied in stages where the stages in the beginning comprise of simpler features whereas the features in the later stages comprise of complex features. In case a feature fails on a particular window in an image, the algorithm concludes that no facial features are present in the image and the remaining features are not run on the image. The algorithm progresses in the aforementioned manner eventually detecting a face.

*Linear Binary Pattern Histogram (LBPH):* LBPH is a technique that builds on top of the technique of Linear Binary Patterns (LBP). LBP labels the pixels of an image by thresholding the neighborhood of each pixel and considers the result as a binary number. The idea behind LBPH is that LBP when combined with histograms represents facial images as a simple data vector. The first step in LBPH is the creation of a master list with images that the algorithm needs to recognize. A unique ID is needed to be assigned to all the images that belong to a certain individual. A window of a certain dimension (say 3 x 3) is extracted from the grayscale representation of the original image. The window may be represented as a matrix with each cell containing the intensity of a pixel. The central value in the matrix is considered the threshold. For each of the neighbors of the threshold, a value of 1 is assigned if the value is greater than or equal to the threshold, 0 otherwise, resulting in a matrix containing only binary values. The binary values in the matrix are concatenated line-by-line in clockwise direction and converted to a decimal value. The decimal value is assigned to the location of the threshold in the matrix. This process when repeated by sliding the window across the image results in a new image that better represents the characteristics of the original image. Grid parameters i.e. the predefined cells in the horizontal and vertical direction (usually 8 x 8) are used to divide the image into multiple grids and a histogram is generated for each grid in the image. Each of the histograms is then concatenated resulting in a large histogram that represents the characteristics of the original image. In order to recognize a face, the LBPH algorithm is run and the resulting histogram is compared against the histograms of each of the images in the master list to identify the best match.

### IV. ARCHITECTURE AND IMPLEMENTATION

#### A. Architecture

Figure 1 shows the modularized architecture of PiBase. The system comprises of two communication channels established over the internet – one between the Android smartphone and Google Firebase, another between Google Firebase and Raspberry Pi.

*Dataflow from Android App to Raspberry Pi:* In the channel between the Android smartphone and Google Firebase, the Android smartphone leverages the authentication system provided by Google Firebase to register and authenticate users on PiBase. Besides, the user is allowed to upload images of people that PiBase may perceive of as non-intruders. The images are stored in the Storage of Google Firebase and a corresponding record is inserted in the real-time database. On the second channel between Google Firebase and Raspberry Pi, the Raspberry Pi pulls down the images uploaded by the Android app. The images received by the Raspberry Pi are used to train the face recognition algorithm.

*Dataflow from Raspberry Pi to Android App:* Upon the detection of an intruder, the Raspberry Pi uploads the image to the real-time database of Google Firebase. Upon recognizing a write to the real-time database, Google Firebase triggers a message to each of the users subscribed to receive notifications of the intrusion.

## B. Implementation

As evident from the architecture, the system consists of three major components – Android smartphone, Google Firebase and Raspberry Pi. The primary use case of the application is to detect intruders and notify users. The implementation pertinent to each of the components is as described below.

   a. Android Application

The Android application is implemented in Java on Android Studio integrated development environment (IDE). The flow of the application on the smartphone is as shown in Figure 2. The app is connected to Google Firebase using the Google Firebase Console since the app relies on Firebase's services for authentication and database. Firstly, the user is required to register with PiBase where the user's name, email and password of their choice are collected as shown in Figure 3(a). The credentials are sent to Google Firebase which creates a new account linked to the credentials. The information is stored as part of the Firebase project and in the event multiple apps are connected to a single Firebase project, the credentials may be used to authenticate the user on any app of the project. Upon registration, the user may use the registration credentials to login to the application. Upon a successful login, the user is presented with options to upload name and images of non-intruders or view past intrusions. Multiple images of each of the non-intruders are desired to aid in detection. The images and names of each of the non-intruders are added to the upload screen as shown in Figure 3(b). Clicking on the 'Choose Image' button lets the user select images from the smartphone's filesystem. The user may view the next or previous images by clicking on the 'Next Image' or 'Previous Image' button respectively. The uploaded images are sent to storage on Google Firebase and stored in a folder with the name provided on the upload screen. Besides, a record containing the folder name of the inserted images and timestamp is inserted in the real-time database which the Raspberry Pi later uses to check if new non-intruders are added to the database. In order to view past intrusions, the user may pick a start date and end date. The intrusions within the date boundary are displayed as a list on the app. On the flip side, Google Firebase sends notifications to the app anytime an intrusion is detected by Raspberry Pi. The notifications are visible on Android's notification tray as shown in Figure 3(c). The user may click on the notification to view in the app.

   b. Google Firebase

The backend services for the application are provided by Google Firebase, a cloud environment. The platform Admin SDK provided by the platform is used to perform user authentication through an Android app. The *signInWithEmailAndPassword()* method is used to pass the email and password credentials of the user from Android. The method has callback listeners which indicate the success or failure of the login. Another key feature of Firebase's that is leveraged is Firebase Cloud Messaging. As mentioned in section III, Google Firebase is able to send notifications to users. Since all of the apps connected to a Firebase project are required to be notified right away upon the detection of an intrusion which involves a write to the real-time database, Firebase Cloud Functions are used. Cloud Functions is a serverless framework that executes a predefined set of instructions or code in response to an event. A cloud function that is triggered in response to a write/ update on the database is setup on Google Firebase. Figure 4 shows a snippet of the trigger that sends a data message to the Android app upon receiving an intrusion image from the Raspberry Pi on a topic named 'rpi_security'.

   c. Raspberry Pi

The Raspberry Pi 3.0 is equipped with the Pi Camera module and PIR motion sensor. Figure 5 shows the circuit diagram of the PIR motion sensor and Pi Camera module connected to the Raspberry Pi. The camera is inserted into the camera module port of the Raspberry Pi. The camera's blue color strip needs to face the ethernet port. For the PIR motion sensor, the voltage pin may be inserted into any one of the DC voltage (5V) pins of the Raspberry Pi, the ground (GND) pin may be connected to any one of the GND pins on the Raspberry Pi and Digital OUT pin to one of the GPIO input pins on the Raspberry Pi. The PIR motion sensor and camera act in tandem to detect intruders. Since the Raspberry Pi is resource constrained, running both PIR motion sensor and camera at all times hogs the resources for the Raspberry Pi to perform other tasks. In order to prevent it, only the PIR motion sensor is in operation at all times. The camera is only activated upon a detection of motion by the PIR sensor. The dataflow of PiBase on the Raspberry Pi is as shown in Figure 6. Since the goal is to detect and recognize an intruder, of the essence is an image where the face is clearly visible. However, the person in the vicinity of the camera moves and does not always face the camera head-on. So, face detection and recognition are not an easy task. In order to best achieve an image with a clear face, the camera is setup to capture ten images with an interval of two seconds between each image in hopes that at least one of the images shows the face clearly. Practically, it helps to have an enticing object below the Raspberry Pi setup as bait to attract the intruder towards the camera. Each of the captured images is run through the Haar Cascades face detection algorithm. In the event a face is detected, the face is put through the LBPH face recognition algorithm to identify the face. The LBPH algorithm is required to be trained on the non-intruder image dataset before it can be put to the task of facial recognition. In order to do so, the Raspberry Pi pulls down the images from Google Firebase once a day. Since the Android app inserts a record containing the folder name and timestamp in the real-time database, an automation script written in Python and triggered daily on the Raspberry Pi retrieves non-intruder images based on the records inserted during the past 24-hour time frame. Besides, the automation script initiates training of the LBPH algorithm on the images. Training the LBPH algorithm right after the Android app uploads a non-intruder is desired. However, training the LBPH algorithm consumes resources on the resource-constrained Raspberry Pi and training the algorithm multiple times a day might hinder the performance of the device. One of the features of LBPH is that a confidence score is output along with the prediction. Unlike in the conventional sense, a low confidence score is desired. In case LBPH does not match the image against any of the images in the master list, the detected image is uploaded from the Raspberry Pi to the real-time database in Google Firebase.

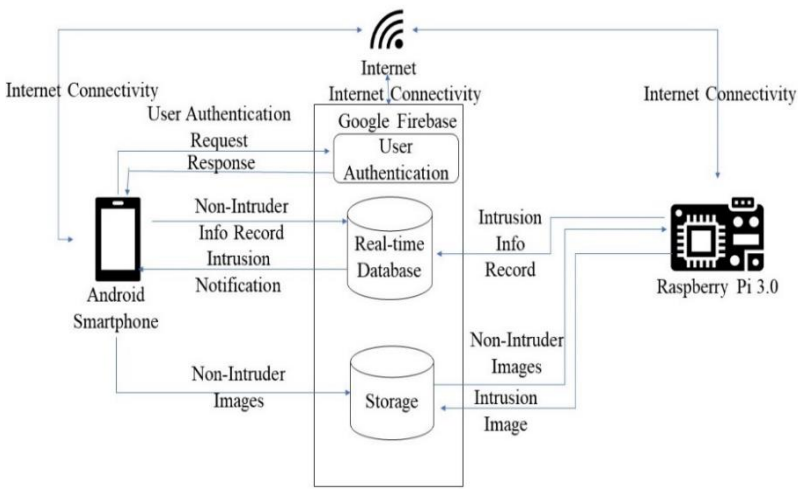

*Figure 1: PiBase Architecture*

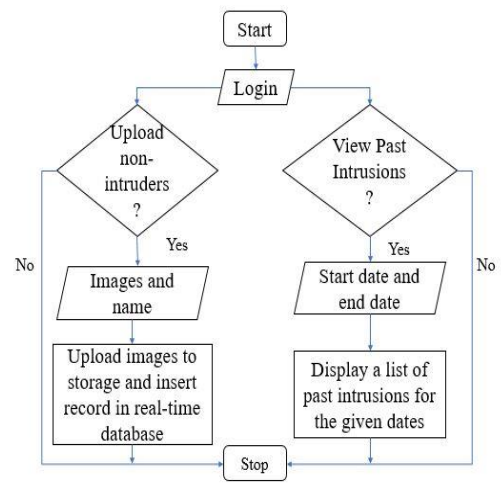

*Figure 2: Android Smartphone Dataflow*

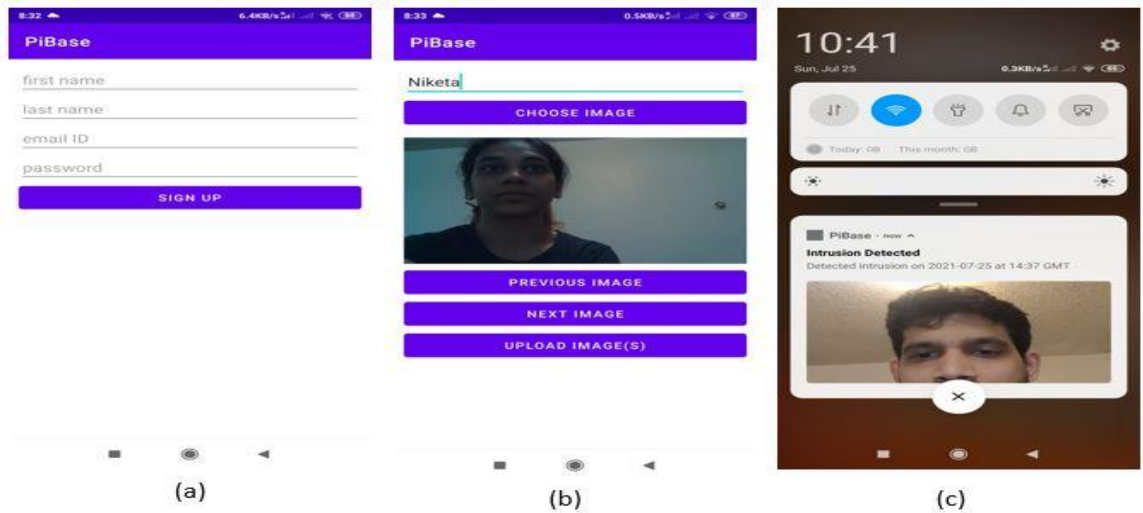

*Figure 3: (a) User Registration Screen; (b) Non-Intruder Image Upload Screen; (c) Intrusion Notification from Google Firebase*

## C. Experiment

The Raspberry Pi is installed inside a home environment and tested on a real-world dataset containing a master list of 23 people with fifteen males and eight females. Six images of each people's faces are captured with the head in different orientations. When put to test wherein each of the 23 people is made to take a stroll inside the room four to five times, the Haar Cascades face detection algorithm detects 106 faces over a total of 112 trials. The metrics of Precision and Recall are used to evaluate the face detector. Precision refers to the number of positive correctly identified by the classifier and is given by $true\ positives / (true\ positives + false\ positives)$ whereas Recall refers to the number of positives identified of all the positives in the dataset and is

```
exports.sendAdminNotification = functions.database.ref("/Users/{pushId}")
    .onWrite((event) =>{
    const payload= {notification: {
      title: "Intrusion Detected",
      body: "Detected intrusion on " + date + " at " + time + " GMT"
    },
    data: {
      imageUrl: image
    },
  };
    return admin.messaging().sendToTopic("rpi_security", payload)
      .then(function(response) {
        console.log("Notification sent successfully:", response);
      }).catch(function(error) {
        console.log("Notification sent failed:", error);
      });
  });
```

*Figure 4: Firebase Trigger on Real-time Database*

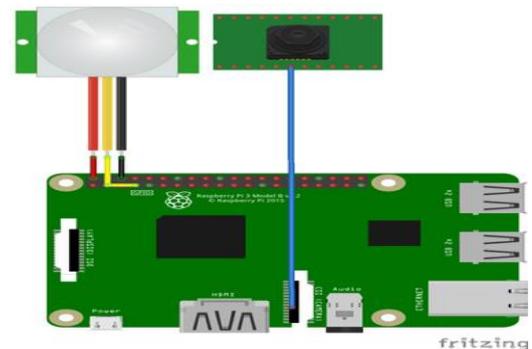

*Figure 5: Raspberry Pi Connected to Pi Camera and PIR Motion Sensor [17]*

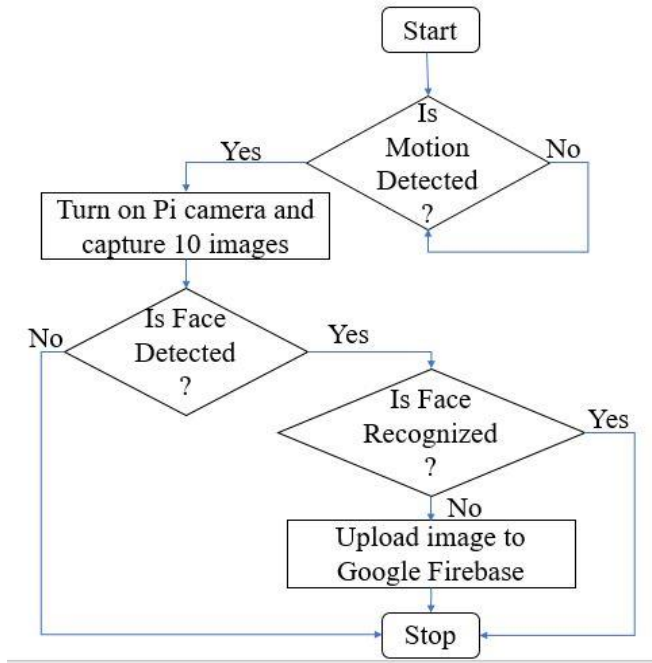

*Figure 6: Raspberry Pi 3.0 Dataflow*

given by $true\ positives/(true\ positives + false\ negatives)$. The face detector exhibits a Precision of 1.0 since no false positives are detected. The recall is 94.64%. The detected images are run through the LBPH face recognizer. The face recognizer does not misclassify an individual i.e. no false positives. Hence, the Precision is 1.0. The Recall at different thresholds of confidence scores is as shown in Table 1.

*Table 1: Face Recognition Recall of LBPH*

Figure 7 shows the workflow of PiBase's image capture, face detection and face recognition on an individual in the vicinity of the camera. Inferring from the results in Table 1, the percentage of recognition improves as the bounds are loosened. While thresholds are user defined and may be loosened at will, it is imperative that the algorithm learns to recognize the faces with a high level of certainty. From the experiments, the observation is that the face recognizer works best when the entities in question have distinct features. Facial hair seems to have an impact since the recognition confidence scores for males with a peculiar beard pattern are lower (better) than males without beards.

## V. FUTURE WORK AND CONCLUSION

While face detection and recognition are performed decently well, there is room for improvement. In future, the idea is experiment with pre-trained and home-grown architectures of deep neural networks meant for resource constrained devices such as the Raspberry Pi. Additional functionality such as an alarm system that may be triggered from the Android app is in consideration.

Overall, PiBase is a low-cost security system implemented on Android and Raspberry Pi with Google Firebase acting as the backend. The system demonstrates the feasibility of IoT and mobile applications to security systems. While the proposed system is applied in terms of home security, the idea of face detection and recognition may be applied in any field where the need for authorized access persists.

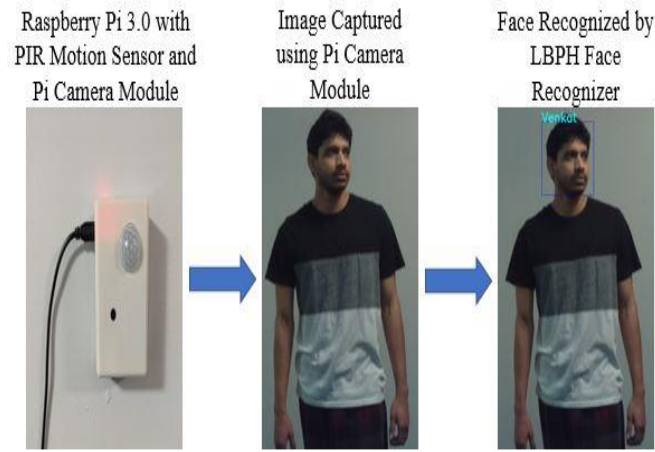

*Figure 7: PiBase Image Capture and Face Recognition*